\documentclass[12pt]{article}
\usepackage{a4}
\begin{document}
\renewcommand\title[1]{\begin{center}\Large\bf#1\end{center}}
\renewcommand\author[1]{\begin{center}\large\bf#1\end{center}}
\newcommand\aff[1]{\begin{flushleft}\normalsize#1\end{flushleft}}
\newcommand\email[1]{\footnote{#1}}
\title{Predictions for $\mu\rightarrow e\gamma$ in SUSY from non trivial Quark-Lepton complementarity and flavor symmetry}
\author{Marco Picariello
\email{Marco.Picariello@le.infn.it}}
\aff{
Dipartimento di Fisica - Universit\`a di Lecce and INFN - Lecce, Italia\\
\centerline{\rm and}
CFTP - Dep. de F\'{\i}sica - Instituto Superior T\'{e}cnico - Lisboa, Portugal
}

\begin{abstract}
{We compute the effect of non diagonal neutrino
mass in $l_i\rightarrow l_j\gamma$
in SUSY theories with non trivial Quark-Lepton
complementarity and a flavor symmetry.
The correlation matrix $V_M=U_{CKM}U_{PMNS}$ is such that its
$(1,3)$ entry, as preferred by the present experimental data, is zero.
We do not assume that $V_M$ is bimaximal.
\\
Quark-Lepton complementarity and the flavor symmetry strongly constrain
the theory and we obtain a clear prediction for the contribution to
$\mu\rightarrow e\gamma$ and the $\tau$ decays
$\tau\rightarrow e\gamma$ and $\tau\rightarrow \mu\gamma$.
If the Dirac neutrino Yukawa couplings are degenerate but the
low energy neutrino masses are not degenerate, then the
lepton decays are related among them by the $V_M$ entries.
On the other hand, if the Dirac neutrino Yukawa couplings are
hierarchical or the low energy neutrino masses are degenerate,
then the prediction for the lepton decays comes from the $U_{CKM}$
hierarchy.
}
\end{abstract}
{{\bf keywords}: Neutrino mass matrices, SUSY, Quark-Lepton complementarity, flavor symmetry, $\mu\rightarrow e\gamma$}

%\preprint{DRAFT}

\section{Introduction}
The present experimental situation is such that we are
very close to obtain a theory of flavor that is able
to explain in a clear way all the Standard Model masses and
mixing.
The last but not least experimental ingredient has been the
neutrino data and the determination of $\Delta m_{12}^2$,
$|\Delta m_{23}^2|$, $\theta_{12}$ and $\theta_{23}$.
From all these results we are able to extract strong
constraints on the flavor structure of the SM.
In particular the neutrino data were determinant to
clarify the role of the discrete symmetry in flavor physics.

The disparity that nature indicates between quark and lepton mixing
angles has been viewed in terms of a 'Quark-Lepton complementarity' (QLC)
\cite{Raidal:2004iw,Minakata:2004xt} which can be expressed in the relations
\begin{eqnarray}
\theta_{12}^{PMNS}+\theta_{12}^{CKM}\simeq 45^\circ\,;
\quad\quad
 \theta_{23}^{PMNS}+\theta_{23}^{CKM}\simeq 45^\circ\,.
\end{eqnarray}
Despite the naive relations between the PMNS and CKM angles, a detailed analysis shows that the correlation matrix $V_M=U_{CKM}U_{PMNS}$
is phenomenologically compatible with a tribimaximal pattern, and only marginally with a bimaximal pattern.
Future experiments on neutrino physics, and in particular
in the determination of $\theta_{23}$ and the CP violating parameter
$J$, will be able to better clarify if a trivial Quark-Lepton
complementarity, i.e. $V_M$ bimaximal, is ruled out in favor of
a non trivial Quark-Lepton complementarity,
i.e. $V_M$ tribimaximal or even more structured \cite{Picariello:2006sp}.
From present experimental evidences a non trivial Quark-Lepton
complementarity arises \cite{Chauhan:2006im}.
Moreover the clear non trivial structure of $V_M$ 
and the strong indication of gauge
coupling unification allow us to obtain in a straightforward way
constraints on the high energy spectrum too.
Within this framework we get information
about flavor physics from the correlation matrix $V_M$ too.
It is very impressive that for some discrete flavor symmetries
such as $A_4$ dynamically broken into $Z_3$ 
\cite{Morisi:2007ft,Altarelli:2005yp,Bazzocchi:2007au}
or $S_3$ softly broken into $S_2$ 
\cite{Morisi:2005fy,added,Caravaglios:2005gw} the tribimaximal structure
appears in a natural way.

In supergravity theories if the effective
Lagrangian is defined at a scale higher than the Grand Unification scale,
the matter fields
have to respect the underlying gauge and flavor symmetry.
Hence, we expect quark-lepton correlations among entries of the sfermion mass matrices.
In other words, the quark-lepton unification seeps also into the SUSY
breaking soft sector \cite{Ciuchini:2007ha}.
In general we do not get strongly renormalization effects
on flavor violating quantities from the heavy neutrino
scale to the electroweak scale because of the absence 
of flavor violation. In fact the remaining flavor
violation related to the low energy neutrino sector
gives a negligible contribution with the exception of
the case with highly degenerate neutrinos and $\tan\beta>40$
\cite{Antusch:2005gp,Schmidt:2006rb}.

In this work we compute the effect of non diagonal neutrino
mass in $l_i\rightarrow l_j\gamma$ 
in SUSY theories with non trivial Quark-Lepton
complementarity and flavor symmetry.
In comparison with previous works (i.e. \cite{Hochmuth:2006xn,Cheung:2005gq}),
where a bimaximal $V_M$ matrix is assumed, in the present work
the correlation matrix $V_M=U_{CKM}U_{PMNS}$ is such that its
$(1,3)$ entry, as preferred by experimental data, is zero.
All the other entries are assumed to vary as allowed by the
experimental data \cite{Picariello:2006sp,Chauhan:2006im}.
Nevertheless We obtain a clear prediction for the contribution to
$l_i\rightarrow l_j\gamma$.
By using the non trivial Quark-Lepton complementarity, flavor symmetry,
and the see-saw mechanism we will compute the explicit 
spectrum of the heavy neutrinos. This will allow us to
investigate the relevance of the form of $V_M$ in
$l_i\rightarrow l_j\gamma$.
There are three cases. They depend on the spectrum of
the Dirac neutrino mass matrix and the low energy neutrinos.
We may have:
1) hierarchical Dirac neutrino eigenvalues (in this case we
have very hierarchical right-handed neutrino masses);
2) degenerate Dirac neutrino eigenvalues, with non degenerate
low energy neutrino masses (in this case the
hierarchy of the right-handed neutrino masses is close to the
one of the low energy spectrum);
3) degenerate Dirac neutrino eigenvalues and low energy neutrino
spectrum (that implies right-handed neutrinos close to degenerate).
For each of these cases we have different contributions to
$l_i\rightarrow l_j\gamma$.
We will show that only when Dirac
neutrino eigenvalues are degenerate and 
low energy neutrino masses are not degenerate,
the explicit form of $V_M$ plays an important role.

The plan of the work is the following.
In Sec. {\bf\ref{sec:notation}} we explain our notations and
clarify the meaning of the correlation matrix $V_M$ in flavor theories.
In Sec. {\bf\ref{sec:Obs}} we introduce the relation between
$l_i\rightarrow l_j\gamma$ and the Dirac neutrino matrix.
In Sec. {\bf\ref{sec:MD}} we relate the Dirac neutrino Yukawa coupling
to the $CKM$ mixing matrix by using the non trivial Quark-Lepton
complementarity and flavor symmetry. Then we compute the heavy neutrino spectrum.
In Sec. {\bf\ref{sec:muegamma}} we compute the value of
the contribution to the $l_i\rightarrow l_j\gamma$ processes from
a non diagonal Dirac neutrino Yukawa coupling.
Finally in Sec. {\bf\ref{sec:conclusion}} we report our conclusions.

\section{Notations}\label{sec:notation}
In this section we explain the relation
between the product $V_M=U_{CKM}U_{PMNS}$ and the diagonalization
of the right-handed neutrino mass.
\subsection{$V_M$ in theories with see-saw of type I} 
Let's fix the notations in the lepton sector.
Let $Y_l$ be the Yukawa matrix for charged leptons.
It can be diagonalized by
\begin{eqnarray}\label{eq:UlVl}
Y_l&=&U_l Y_l^\Delta V_l^\dagger\,.
\end{eqnarray}
Let $M_R$ be the Majorana mass matrix for the right-handed neutrino
and $M_{D}$ the Dirac mass matrix.
Under the assumption that the low energy neutrino masses are given by
the see-saw of Type I we have that the light neutrino mass matrix
is given by
\begin{equation}\label{eq:Mnu}
M_\nu = M_{D} \frac{1}{M_R} M_{D}^T\,.
\end{equation}
Let us introduce $U_0$ from the diagonalization of the Dirac mass
matrix
\begin{eqnarray}\label{eq:Dirac}
M_{D} &=& U_0 M_{D}^\Delta V_0^\dagger\,,
\end{eqnarray}
then we define $V_M$ by the diagonalization of the light neutrino mass
\begin{eqnarray}\label{eq:light}
M_\nu&=&U_\nu M_\nu^\Delta U_\nu^T
\nonumber\\&=& U_0 V_M M_\nu^\Delta (V_M)^T U_0^T\,,
\end{eqnarray}
with the constraint that $U_0 V_M$ is an unitary matrix.
Finally the lepton mixing matrix is
\begin{eqnarray}\label{eq:PMNS}
U_{PMNS} = U_l^\dagger U_\nu = U_l^\dagger U_0 V_M \,.
\end{eqnarray}
Let us introduce the following symmetric complex matrix ${\cal C}$
\begin{eqnarray}\label{eq:C}
{\cal C}&=&M_D^\Delta V_0^\dagger \frac{1}{M_R} V_0^\star M_D^\Delta\,,
\end{eqnarray}
where $V_0$ is the mixing matrix that diagonalizes on the right the
Dirac neutrino mass matrix.
From eqs. (\ref{eq:Dirac}-\ref{eq:light}) we see that 
the inverse of $V_M$ diagonalizes the symmetric matrix $\cal C$,
in fact we have
\begin{eqnarray}
V_M M_\nu^\Delta V_M^T &=& {\cal C}\,.
\end{eqnarray}
\subsection{Flavor symmetry implies $V_M$ as correlation matrix}
\label{sec:2.2}
In the quark sector we introduce $Y_u$ and $Y_d$
to be the Yukawa matrices for up and
down sectors.
They can be diagonalized by
\begin{eqnarray}
Y_u=U_u Y_u^\Delta V_u^\dagger&\mbox{and}&Y_d=U_d Y_d^\Delta V_d^\dagger\,,
\end{eqnarray}
where the $Y^\Delta$ are diagonal and the $U$s and $V$s are unitary matrices.
\\
Then the quark mixing matrix is given by
\begin{eqnarray}\label{CKM}
U_{CKM}&=&U_u^\dagger U_d\,.
\end{eqnarray}
To relate the $U_{CKM}$ with the $U_{PMNS}$ normally one makes use
of GUT models, such us generic $SO(10)$ or $E_6$, where 
there are some natural Yukawa
unifications. In fact these cases give an interesting relation
between the $U_{CKM}$ quark mixing matrix, the $U_{PMNS}$ lepton mixing matrix
and $V_M$ obtained from eq. (\ref{eq:C}).
The mixing matrix $V_M$ turns out to be the correlation matrix defined
in eq. (\ref{eq:fund}).
The reason for it is that in $SO(10)$ or $E_6$ one has intriguing
relations between the Yukawa couplings of the quark sector and that
of the lepton sector.
For instance, in minimal renormalizable $SO(10)$ with Higgs in the
$\bf 10$, $\bf 126$, and $\bf 120$, we can have $Y_l \approx Y_d^T$.

However this feature is much more general and may depend on
the flavor symmetry instead of the gauge grand unification.
The presence of a flavor symmetry usually implies the structure of the
Yukawa matrices and the equivalent entries of
$Y_l$ and $Y_d$ are of the same order of magnitude.
We conclude that, as long as the flavor symmetry fully constraints the
mixing matrices that diagonalize the Yukawa matrices, we have $U_l \simeq V_d^\star$.
Notice that if there is a flavor symmetry that
constrains the Yukawa couplings in such a way that the diagonalizing
unitary matrices are fixed, then the entries of $Y_l$ can still be
very different from the entries of $Y_d^T$. However both Yukawa
matrices are diagonalized by the same mixing matrices.
This is exactly the case in the presence of an
$A_4$ discrete flavor symmetry dinamically broken into $Z_3$
\cite{Morisi:2007ft,Altarelli:2005yp,Bazzocchi:2007au}
and can be partially true in the case of $S_3$ softly broken into
$S_2$ \cite{Morisi:2005fy,Caravaglios:2005gw}.
\\
From eq. (\ref{eq:PMNS}) we get
$$ U_{PMNS} \simeq V_d^T U_0 V_M\,.$$
If we denote by $Y_\nu$ the Yukawa coupling that generates the Dirac neutrino
mass matrix $M_D$, we have also the relation
\begin{eqnarray}
Y_\nu \approx Y_u^T &\rightarrow& U_0 \simeq V_u^\star\,.
\end{eqnarray}
This relation, together with the previous one, implies
$$U_{PMNS} \simeq V_d^T V_u^\star V_M\,.$$
If the Yukawa matrices are diagonalized by a similar matrix on the
left and on the right, for example in minimal renormalizable
$SO(10)$ with only small contributions from the antisymmetric
representations such as ${\bf 120}$ or more important in models
where the diagonalization is strongly constrained by the flavor
symmetry, the previous relationship translates
into a relation between $U_{PMNS}$, $U_{CKM}$ and $V_M$.
In fact we have
\begin{eqnarray}\label{VM}
Y_u   \simeq Y_u^T \rightarrow V_u^\star = U_u &\mbox{and}&
Y_d   \simeq Y_d^T \rightarrow V_d^\star = U_d\,. \nonumber
\end{eqnarray}
The first relation tells us that 
$$U_{PMNS} = V_d^T U_u V_M\,.$$
Finally, using the second relation in eq. (\ref{VM})
and the definition of the CKM mixing
matrix of eq. (\ref{CKM}) we get
\begin{equation}\label{eq:fund}
V_M=U_{CKM}\cdot \Omega \cdot U_{PMNS}\,,
\end{equation}
where we introduced the matrix
\begin{eqnarray}\label{Omega}
\Omega={\rm diag}(e^{i \omega_1},e^{i \omega_2},e^{i\omega_3})\,
\end{eqnarray}
to allow us to write the $CKM$ and $PMNS$ matrices in their standard
form (i.e. three rotation angles and one phase for the $CKM$ and the
equivalent for the $PMNS$) and to take into account the
phase mismatching between quarks and leptons.
The form of $V_M$ can be obtained under some assumptions about the
flavor structure of the theory.
Some flavor models give for example a correlation $V_M$ with $(V_M)_{13}=0$.
As a consequence of the from of the non trivial Quark-Lepton 
complementarity there are some predictions from the model,
such as for $\theta_{13}^{PMNS}$ from \cite{Chauhan:2006im}
and the correlations between CP violating phases and the mixing angle
$\theta_{12}$ of \cite{Picariello:2006sp}.

\section{The observables}\label{sec:Obs}
As explained in the introduction, in this work we are interested
in extracting informations from non trivial
quark-lepton complementarity and flavor symmetry about
the $l_i\rightarrow l_j \gamma$ decays.
We report here the usual formula obtained in the literature
on these processes.
It is obtained in the weak eigenstate neutrino base, 
where charged lepton and Majorana right-handed neutrino
mass matrices and weak interactions are diagonal.
These processes depends on $\tilde M_D$, the Dirac neutrino mass
in the weak base.

\subsection{$l_i\rightarrow l_j\gamma$}
The contribution at first order approximation
to the process $l_i\rightarrow l_j\gamma$ in SUSY
models is given by
\begin{eqnarray}\label{eq:BR}
BR(l_i\rightarrow l_j\gamma)
     &\propto&
\frac{\Gamma(l_i\rightarrow e\nu\nu)}{\Gamma(l_i)}
\frac{\alpha^3}{G_f m_s^8 v_u^4}\tan^2\beta
\left(\frac{3m_0+A_0}{8\pi^2}\right)^2
\left|\left(\tilde M_D L \tilde M_D^\dagger\right)_{ij}\right|^2
\end{eqnarray}
where $m_0$ is the universal scalar mass,
$A_0$ is the universal trilinear coupling parameter,
$\tan\beta$ is the ratio of the vacuum
expectation values of the up and down Higgs doublets,
and $m_s$ is a typical mass of superparticles
with \cite{Petcov:2003zb}
$ m^8_s\approx 0.5 m_0^2 M^2_{1/2}(m^2_0+ 0.6 M^2_{1/2})^2$,
where $M_{1/2}$ is the gaugino mass.
The matrix $L_{ij}={\bf 1}_{ij}\log M_x/M_i$ takes into account the
RGE effects on the Majorana right-handed neutrino masses.
In fact the eq. (\ref{eq:BR}) is computed in the base where the
Yukawa of the charged lepton and the Majorana neutrino mass
are diagonal.
Eq. (\ref{eq:BR}) is valid in the base where right-handed Majorana
neutrino mass matrix, charged lepton mass matrix and weak
gauge interactions are diagonal.
The experimental limit for the branching ratio of
$\mu\rightarrow e\gamma$ is $1.2\times 10^{-11}$ at $90\%$ of confidence
level \cite{Brooks:1999pu} and it could go
down to $10^{-14}$ as proposed by MEG collaboration.

\section{$\tilde M_D$ from non trivial Quark-Lepton complementarity and flavor symmetry}
\label{sec:MD}
Let us investigate the value of Dirac neutrino mass matrix $\tilde M_D$
in the base where right-handed Majorana neutrino mass matrix,
 charged leptons mass matrix and weak gauge interactions
 are diagonal.
The part of the Standard Model Lagrangian containing the leptons is
\begin{eqnarray}
{\cal L}&=&\bar\nu_L Y_D \nu_R H 
 + \nu_R^T C M_R \nu_R + \bar l_L Y_l l_R H +
   \bar\nu_L W\!\!\!\!\!/\ \ l_L\,.
\end{eqnarray}
We want to redefine the fields in such a way that the only source
of flavor violation is in the Dirac neutrino Yukawa coupling.
We introduce the following definitions
\begin{eqnarray}
l_R^\prime  =V_l^\dagger\ l_R  \,,  \quad\quad
\nu_R^\prime=V_R^T\ \nu_R      \,,  \quad\quad
l_L^\prime  =U_l^\dagger\ l_L  \,,  \quad\quad
\nu_L^\prime=U_l^\dagger\ \nu_L\,,
\end{eqnarray}
where the unitary matrices $V_l$, $U_l$
are defined in eqs. (\ref{eq:UlVl}).
The unitary matrix $V_R$ is defined by the diagonalization of $M_R$
\begin{eqnarray}\label{eq:MRdiag}
V_R M_R^\Delta V_R^T = M_R\,.
\end{eqnarray}
Consequently we have
\begin{eqnarray}
l_R    = V_l\ l_R^\prime    \,,\ 
\nu_R  = V_R^*\ \nu_R^\prime\,,\ 
l_L    = U_l\ l_L^\prime  
&
\mbox{and}
&
\nu_R^T   = (\nu_R^\prime)^T\ V_R^\dagger \,, \quad
\bar l_L  =\bar l_L^\prime\   U_l^\dagger \,, \quad
\bar \nu_L=\bar \nu_L^\prime\ U_l^\dagger \,.
\nonumber
\end{eqnarray}
In this primed base
%, called the weak eigenstates base for the neutrino? NO
we get
\begin{eqnarray}
{\cal L}&=&\bar\nu_L^\prime U_l^\dagger M_D V_R^*\nu_R^\prime
 + (\nu_R^\prime)^T C M_R^\Delta \nu_R^\prime + \bar l_L^\prime
  M_l^\Delta l_R^\prime +
   \bar\nu_L^\prime W\!\!\!\!\!\!/\ \ l_L^\prime
\end{eqnarray}
and we define
\begin{eqnarray}
\tilde M_D&=&U_l^\dagger M_D V_R^\star\,.
\end{eqnarray}
We want now to relate this $\tilde M_D$ matrix to the $CKM$ mixing
matrix by using the non trivial Quark-Lepton complementarity and flavor symmetry.
First of all we rewrite this matrix as
\begin{eqnarray}\label{eq:MtLM}
\tilde M_D &=&U_l^\dagger M_D V_R^\star
\nonumber\\&=&
U_l^\dagger U_0 M_D^\Delta V_0^\dagger V_R^\star\,.
\end{eqnarray}
Then we notice that the matrix $V_0^\dagger V_R^\star$ is related
via the ${\cal C}$ matrix to the diagonal low energy neutrino
mass matrix $m_{low}^\Delta$ and to $V_M$.
In fact we have
\begin{eqnarray}
V_M m_{low}^\Delta V_M^T &=& {\cal C}
\nonumber\\&=&
M_D^\Delta V_0^\dagger \frac{1}{M_R} V_0^\star M_D^\Delta
\nonumber\\&=&\label{eq:VM}
M_D^\Delta V_0^\dagger V_R^\star
\frac{1}{M_R^\Delta} V_R^T V_0^\star M_D^\Delta
\end{eqnarray}
where we used the inverse of eq. (\ref{eq:MRdiag})
\begin{eqnarray}
V_R^\star \frac{1}{M_R^\Delta} V_R^\dagger = \frac{1}{M_R}\,.
\end{eqnarray}
We multiply on the left and on the right both sides of eq. (\ref{eq:VM})
by $1/M_D^\Delta$ and we get
\begin{eqnarray}\label{eq:V0VR}
V_0^\dagger V_R^\star\frac{1}{M_R^\Delta} V_R^T V_0^\star
&=&
\frac{1}{M_D^\Delta} V_M m_{low}^\Delta V_M^T \frac{1}{M_D^\Delta}\,.
\end{eqnarray}
If one uses the method of \cite{Casas:2001sr} one can extract the
matrix $V_0^\dagger V_R^\star$ by making the {\em square root} of the
matrices in eq. (\ref{eq:V0VR}). One has
\begin{eqnarray}
V_0^\dagger V_R^\star\sqrt{\frac{1}{M_R^\Delta}}
&=&
\frac{1}{M_D^\Delta} V_M \sqrt{m_{low}^\Delta} R^T\,,
\end{eqnarray}
where $R$ is a complex orthogonal matrix such that $R^T R={\bf 1}$,
and one obtains
\begin{eqnarray}
V_0^\dagger V_R^\star
&=&
\frac{1}{M_D^\Delta} V_M \sqrt{m_{low}^\Delta} R^T
\sqrt{M_R^\Delta}\,.
\end{eqnarray}
Finally one concludes that
\begin{eqnarray}
\tilde M_D
&=&U_l^\dagger U_0 M_D^\Delta
\frac{1}{M_D^\Delta} V_M \sqrt{m_{low}^\Delta} R^T\sqrt{M_R^\Delta}
\\\label{eq:Ibarra}
&=&
 U_{PMNS} \sqrt{m_{low}^\Delta} R^T\sqrt{M_R^\Delta}\,.
\end{eqnarray}
Notice that in eq. (\ref{eq:Ibarra}) does not appear
the matrix $V_M$, and any information from $V_M$ is hidden into the
$R$ matrix. 

In our discussion however eq. (\ref{eq:V0VR}) unequivocally fixes
$V_0^\dagger V_R^\star$ and the $R$ matrix, once we know
the eigenvalues of the Dirac neutrino mass matrix and the low energy
neutrino spectrum.
In fact the $V_M$ matrix is assumed to be known because of
the non trivial Quark-Lepton complementarity.
Once we computed the $V_0^\dagger V_R^\star$ matrix form eq. (\ref{eq:V0VR}),
by using eq. (\ref{eq:MtLM}), we get
\begin{eqnarray}
\tilde M_D &=&
U_l^\dagger U_0 M_D^\Delta V_0^\dagger V_R^\star
\nonumber\\&=&
U_{PMNS} V_M^\dagger M_D^\Delta V_0^\dagger V_R^\star
\nonumber\\
&=&\label{eq:Our}
\Omega^\dagger U_{CKM}^\dagger M_D^\Delta V_0^\dagger V_R^\star\,,
\end{eqnarray}
where in the last line we used the relations in eq.
(\ref{eq:PMNS}) and (\ref{eq:fund}).

\subsection{{\em Full} determination of $V_0^\dagger V_R^\star$
 and $M_R^\Delta$}
Eq. (\ref{eq:Our}) is the equivalent of the general 
eq. (\ref{eq:Ibarra}) in presence of non trivial Quark-Lepton
complementarity and flavor symmetry.
We observe that the main modification is the presence of $U_{CKM}^\dagger$
instead of $U_{PMNS}$ thanks to the fact the these matrices
are related to each other through $V_M$ as shown in eq. (\ref{eq:fund}).
Moreover the $R$ is absent and is substantially substituted by 
the {\em known} $V_0^\dagger V_R^\star$ matrix,
computed with eq. (\ref{eq:V0VR}).
Let us now compute the $V_0^\dagger V_R^\star$ matrix in 
a general scenario.
\\
In the following we use the experimental constraint from
\cite{Chauhan:2006im} that says that $(V_M)_{13}$ is zero.
With this single constraint on $V_M$ we write
\begin{eqnarray}
V_M&=&
\left(
\begin{array}{ccc}
\cos\theta_{12}&\sin\theta_{12}&0\\
-\sin\theta_{12}\cos\theta_{23}&\cos\theta_{12}\cos\theta_{23}&\sin\theta_{23}\\
\sin\theta_{12}\sin\theta_{23}&-\cos\theta_{12}\sin\theta_{23}&\cos\theta_{23}
\end{array}
\right)
\end{eqnarray}
and the allowed ranges for $\theta_{12}^{V_M}$
and $\theta_{23}^{V_M}$ are \cite{Chauhan:2006im}
\begin{eqnarray}
	\tan^2\theta_{12}^{V_M}\in[0.3,1.0]
&\quad\mbox{ and }\quad&
	\tan^2\theta_{23}^{V_M}\in[0.5,1.4]
	\,.
\end{eqnarray}
Let us denote by $m_i$ the complex low energy neutrino masses
obtained after the see-saw ($m_{low}^\Delta=\{m_1,m_2,m_3\}$),
and $M_i$ the eigenvalues of the Dirac neutrino mass matrix
$M_D$ ($M_D^\Delta=\{M_1,M_2,M_3\}$). We have
\def\Completa{
\scriptsize
\begin{eqnarray}
V_M m_{low}^\Delta V_M^T
&=&
\left(
\begin{array}{ccc}
(m_1 \cos^2\theta_{12}+m_2\sin^2\theta_{12})&
-(m_1-m_2)\cos\theta_{12}\cos\theta_{23}\sin\theta_{12}&
(m_1-m_2)\cos\theta_{12}\sin\theta_{12}\sin\theta_{23}
\\
-(m_1-m_2)\cos\theta_{12}\cos\theta_{23}\sin\theta_{12}&
(m_1\cos^2\theta_{12}\sin^2\theta_{12}+
 m_2\cos^2\theta_{12}+
 m_3\sin^2\theta_{23})&
\sin\theta_{23}\cos\theta_{23}(m_3-m_2\cos^2\theta_{12}-m_1\cos^2\theta_{12})
\\
(m_1-m_2)\cos\theta_{12}\sin\theta_{12}\sin\theta_{23}&
\sin\theta_{23}\cos\theta_{23}(m_3-m_2\cos^2\theta_{12}-m_1\cos^2\theta_{12})&
\sin^2\theta_{23}(m_1\sin^2\theta_{12}+
 m_2\cos^2\theta_{12})+
 m_3\cos^2\theta_{23}
\\
\end{array}
\right)
\nonumber
\end{eqnarray}
}
{$V_M m_{low}^\Delta V_M^T$ equal to
\small
\begin{eqnarray}
%V_M m_{low}^\Delta V_M^T
%&=&
\left(
\begin{array}{ccc}
(m_1 c^2_{12}+m_2 s^2_{12})&
-(m_1-m_2)c_{12}c_{23}s_{12}&
(m_1-m_2)c_{12}s_{12}s_{23}
\\
-(m_1-m_2)c_{12}c_{23}s_{12}&
(m_1s^2_{12}c^2_{23}+ m_2c^2_{12}c^2_{23}+ m_3s^2_{23})&
s_{23}c_{23}(m_3-m_2c^2_{12}-m_1s^2_{12})
\\
(m_1-m_2)c_{12}s_{12}s_{23}&
s_{23}c_{23}(m_3-m_2c^2_{12}-m_1s^2_{12})&
s^2_{23}(m_1s^2_{12}+ m_2c^2_{12})+ m_3c^2_{23}
\\
\end{array}
\right)
\nonumber\\
\end{eqnarray}
}
and from eq. (\ref{eq:V0VR}) we get
{
\large
\begin{eqnarray}\label{eq:V0VRfull}
%V_0^\dagger V_R^\star\frac{1}{M_R^\Delta} V_R^\dagger V_0^\star
%&=&
\left(
\begin{array}{ccc}
\frac{(m_1 c^2_{12}+m_2 s^2_{12})}{M_1^2}&
\frac{-(m_1-m_2)c_{12}c_{23}s_{12}}{M_1 M_2}&
\frac{(m_1-m_2)c_{12}s_{12}s_{23}}{M_1 M_3}
\\
\frac{-(m_1-m_2)c_{12}c_{23}s_{12}}{M_1 M_2}&
\frac{(m_1s^2_{12}c^2_{23}+ m_2c^2_{12}c^2_{23}+ m_3s^2_{23})}{M_2^2}&
\frac{s_{23}c_{23}(m_3-m_2c^2_{12}-m_1s^2_{12})}{M_2 M_3}
\\
\frac{(m_1-m_2)c_{12}s_{12}s_{23}}{M_1 M_3}&
\frac{s_{23}c_{23}(m_3-m_2c^2_{12}-m_1s^2_{12})}{M_2 M_3}&
\frac{s^2_{23}(m_1s^2_{12}+ m_2c^2_{12})+ m_3c^2_{23}}{M_3^2}
\\
\end{array}
\right)\,.
\nonumber
\end{eqnarray}
}
\vskip-2truecm 
\begin{eqnarray}
\
\end{eqnarray}
Eq. (\ref{eq:V0VRfull}) is general and must be specified depending
on the explicit form of $V_M$. For example for $V_M$ tribimaximal we get
\begin{eqnarray}
V_0^\dagger V_R^\star\frac{1}{M_R^\Delta} V_R^\dagger V_0^\star
&=&
\left(
\begin{array}{ccc}
\frac{2m_1+m_2}{3 M_1^2} &\frac{m_1-m_2}{3M_1 M_2}
  &\frac{m_1-m_2}{3M_1 M_3}\\
\frac{m_1-m_2}{3 M_1 M_2}&\frac{m_1+2m_2+3m_3}{6M_2^2}
  &\frac{m_1+2m_2-3m_3}{6M_2 M_3}\\
\frac{m_1-m_2}{3M_1 M_3} &\frac{m_1+2m_2-3m_3}{6M_2 M_3}
  &\frac{m_1+2m_2+3m_3}{6M_3^2}\\
\end{array}
\right)
\end{eqnarray}
where we remind the reader that $m_i$ are complex numbers, and their sign
is not defined.

\subsection{Hierarchical $M_D$}\label{sec:hier}
First of all let us investigate the case where the
$M_D$ eigenvalues have a hierarchical structure 
as well as any other Dirac mass matrix $M_u$, $M_d$, $M_l$.
As it is well known in this case the heavy neutrino masses
are very hierarchical and the lighter one is very
light compared to the unification scale.
For example if we take the eigenvalues of the Dirac mass matrix $M_D$
to be $M_3\{\lambda^{2 n},\lambda^{n},1\}$ with $n$ of order 1, we get
\footnote{We neglect here the cases
$m_1\simeq m_2\tan^2\theta_{12}$
and
$m_3 \tan^2\theta_{23}\simeq m_1 m_2/(m_1 \cos\theta_{12}+m_2\sin\theta_{12})$.}
\begin{eqnarray}
\frac{1}{M_R^\Delta}&=&
\left(
\begin{array}{ccc}
m_\alpha/(\lambda^{4n} M_3^2)&\ \ 0\ \ &\ \ 0\ \ \\
0&m_\beta/(\lambda^{2n} M_3^2)&0\\
 0 & 0 &m_\gamma/M_3^2
\end{array}
\right) \left(1+O\left(\lambda\right)\right)
\nonumber\\
V_0^\dagger V_R^\star &=&
\left(
\begin{array}{ccc}
1 - \alpha^2 \lambda^{2n}/2 & \alpha\lambda^n &\beta\lambda^{2n}\\
-\alpha\lambda^n & 1- (\alpha^2+ \gamma^2)\lambda^{2n}&\gamma\lambda^n\\
(-\beta + \alpha\gamma)\lambda^{2n} &-\gamma\lambda^n & 1 - \gamma^2\lambda^{2n}/2
\end{array}
\right)+O\left(\lambda^{3n}\right)
\end{eqnarray}
where
\begin{eqnarray}
m_\alpha &=&m_1\cos^2\theta_{12}+m_2\sin^2\theta_{12}+O\left(\lambda^{2n}\right)
\\
m_\beta &=& \frac{m_1 m_2}{m_\alpha}\cos^2\theta_{23}
+m_3\sin^2\theta_{23}+O\left(\lambda^{2n}\right)
\nonumber\\
m_\gamma&=&
\frac{m_1 m_2 m_3}
{m_\alpha m_\beta}
\nonumber\\
\alpha&=&-\frac{(m_1-m_2)}{2 m_\alpha}\sin(2\theta_{12})\cos\theta_{23}
+O\left(\lambda^{2n}\right)
\nonumber\\
\gamma&=&\frac{m_1 m_2- m_3 m_\alpha}
{2 m_\alpha m_\beta}\sin(2\theta_{23})+O\left(\lambda^{2n}\right)
\nonumber\\
\beta&=&
\frac{(m_1-m_2)}{2 m_\alpha}\sin(2\theta_{12})\sin\theta_{23}
+O\left(\lambda^{2n}\right),.
\end{eqnarray}
The numbers $\alpha$, $\beta$, $\gamma$ are of order 1
but the corresponding angles must be computed up to
order $\lambda^{6n}$ to obtain the
right heavy neutrino masses.
The parameters $m_\alpha$, $m_\beta$, $m_\gamma$ are of order of the low
energy neutrino masses.
Notice that the rotation angles $(1,2)$ and $(2,3)$
in $V_0^\dagger V_R^\star$ are of order $\lambda^n$ while the $(1,3)$ 
angle is of order $\lambda^{2n}$.
\\
We observe that in this scenario, with hierarchical Dirac neutrino eigenvalues,
the result depends on the explicit value of the angle $\theta_{12}^{V_M}$
and $\theta_{23}^{V_M}$ only at higher order in $\lambda$ and via
the value of $m_\alpha$, $m_\beta$, $m_\gamma$. For example,
if the $(2,3)$ angle of $V_M$ is $\pi/4$, i.e. for $V_M$ maximal,
 we obtain
\begin{eqnarray}
m_\alpha &=&m_1\cos^2\theta_{12}+m_2\sin^2\theta_{12}+O\left(\lambda^{2n}\right)
\\
m_\beta &=& \frac{m_1 m_2+ m_3m_\alpha}{2m_\alpha}
+O\left(\lambda^{2n}\right)
\nonumber\\
m_\gamma&=&
\frac{m_1 m_2 m_3}{m_\alpha m_\beta}
\nonumber\\
\alpha&=&-\frac{\sqrt{2}(m_1-m_2)\sin(2\theta_{12})}{4m_\alpha}
 +O\left(\lambda^{2n}\right)
\nonumber\\
\gamma&=&1-\frac{m_\gamma}{m_3}+O\left(\lambda^{2n}\right)
\nonumber\\
\beta&=&-\alpha+O\left(\lambda^{2n}\right)
\end{eqnarray}
and for $V_M$ tribimaximal we get
\begin{eqnarray}
m_\alpha &=&\frac{2m_1+m_2}{3}+O\left(\lambda^{2n}\right)
\\
m_\beta &=& \frac{3 m_1m_2+2 m_1m_3+m_2m_3}{2(2m_1+ m_2)}+O\left(\lambda^{2n}\right)
%\nonumber\\&=&
%\frac{2 m(\sqrt{dm_{13}^2} + m)}{\sqrt{dm_{13}^2} + 2 %m}+O\left(\sqrt{\frac{dm_{12}^2}{dm_{13}^2}}\right)
%+O\left(\lambda^{2n}\right)
\\
m_\gamma&=&
\frac{6 m_1 m_2 m_3}{3m_1m_2+2m_1m_3+m_2m_3}\,.
\end{eqnarray}
For any $V_M$, the heavy neutrino spectrum is hierarchical
with ratios given mainly by 
\begin{equation}\label{eq:hierar}
M^R_1:M^R_2:M^R_3\simeq (M_1)^2:(M_2)^2:(M_3)^2\,.
\end{equation}
In fact on one hand we have that,
for normal low energy neutrino hierarchy
$m_\alpha$ is of order $m_2$, $m_\beta$ is of
order $m_3$, and $m_\gamma$ is of order $m_1$.
Then we obtain
$$|m_\alpha| / \lambda^{4n} >> |m_\beta| / \lambda^{2n} >> |m_\gamma|\,.$$
On the other hand, for inverted low energy neutrino hierarchy
$m_\alpha$ is of order $m_2$,
$m_\beta$ is of order $m_1$ ($\approx m_2$), and $m_\gamma$
is of order $m_3$ ($< m_1,m_2$) and then
$$|m_\alpha| / \lambda^{4n} >> |m_\beta| / \lambda^{2n} >> |m_\gamma|\,.$$
Moreover the mixing matrix $V_0^\dagger V_R^\star$
is close to the identity. 
Notice that the lightest right-handed neutrino
has a mass smaller than $M_{Planck} (M_1/M_3)^2$
if we want the mass of the heaviest
right-handed neutrino to be smaller than $M_{Planck}$.

\subsection{Degenerate $M_D$}\label{sec:MDdeg}
We remind the reader that the fact that the non trivial 
quark-lepton complementarity can come from a flavor
symmetry implies that the Dirac neutrino
may have a different hierarchical structure than the
up sector, as clarified in sec. \ref{sec:2.2}.
For example the same argument applies
to the charged lepton and down sectors, where
we know that the hierarchical structure differs
from each other.
The idea beyond this fact, as explained in Sec.
{\bf\ref{sec:notation}}, is that the quark-lepton
complementarity comes both from an unified gauge
theory and from a flavor theory.
It is supposed that, as the
recent progresses show us
\cite{Picariello:2006sp,Chauhan:2006im,Morisi:2007ft,Altarelli:2005yp,Bazzocchi:2007au,Morisi:2005fy,added,Caravaglios:2005gw,Agarwalla:2006dj},
the nature of the mixing angles and that of the mass 
come from different type of flavor symmetries.
For this reason, the non trivial quark-lepton
complementarity can survive even if there
is no Yukawa matrices unification.
The important point is that the mixing
in the Yukawa are related among them.
In Sec. {\bf\ref{sec:notation}} we assumed
these relations, but from recent literature
about flavor physics we know that this is the case.

\subsubsection{Non degenerate $m_{low}$}
If the Dirac neutrino mass eigenvalues are degenerate
then, from eq. (\ref{eq:V0VR}), we obtain 
\begin{eqnarray}
V_0^\dagger V_R^\star\frac{1}{M_R^\Delta} V_R^\dagger V_0^\star
&\simeq&
V_M \frac{1}{M_D^\Delta} m_{low}^\Delta\frac{1}{M_D^\Delta} V_M^T\,.
\end{eqnarray}
In this case, if the low energy neutrino masses are not degenerate,
$V_0^\dagger V_R^\star$ is close to $V_M$ and
$M_R^\Delta \simeq m_{low}^\Delta/(M_D^\Delta)^2$.
Let us define $\delta M_i=M_3-M_i$.
By performing the full computation up to orders
$(\delta M_i/M_3)^2$, we get
\begin{eqnarray}
\frac{1}{M_R^\Delta}&\simeq&
\left(
\begin{array}{ccc}
m_\alpha/M_3^2&\ \ 0\ \ &\ \ 0\ \ \\
0&m_\beta/M_3^2&0\\
 0 & 0 &m_\gamma/M_3^2
\end{array}
\right)
\nonumber\\
V_0^\dagger V_R^\star &\simeq& V_M
\left(
\begin{array}{ccc}
1 - \alpha^2/2 & \alpha &\beta\\
-\alpha & 1- (\alpha^2+ \gamma^2)&\gamma\\
(-\beta + \alpha\gamma) &-\gamma& 1 - \gamma^2/2
\end{array}
\right)
\equiv  V_M V_\epsilon
\end{eqnarray}
where
\begin{eqnarray}\label{eq:MDdeg}
m_\alpha &\simeq&m_1
\left(1
 -\frac{\delta M_1}{M_3}
    \left(1+\frac{\cos(2\theta_{12})}{2}\right)
 +\frac{\delta M_2}{M_3}
    \left(-\frac{1-\cos(2\theta_{12})}{2}-
          \cos(2\theta_{23})\sin^2\theta_{12}\right)
\right)
\nonumber\\
m_\beta &\simeq&m_2
\left(1
 -\frac{\delta M_1}{M_3}
    \left(1-\frac{\cos(2\theta_{12})}{2}\right)
 +\frac{\delta M_2}{M_3}
    \left(-\frac{1-\cos(2\theta_{12})}{2}-
          \cos(2\theta_{23})\cos^2\theta_{12}\right)
\right)
\nonumber\\
m_\gamma&\simeq&m_3
\left(1
  -\frac{\delta M_2}{M_3}\left(1+\cos(2\theta_{23})\right)
\right)
\nonumber\\
\alpha&\simeq&-\frac{m_1+m_2}{4 (m_1-m_2)}
\frac{2 \delta M_1-\delta M_2-\delta M_2\cos(2\theta_{23})}{ M_3}
\sin(2\theta_{12})
\nonumber\\
\gamma&\simeq&\frac{m_2+m_3}{2 (m_2-m_3)}
\frac{\delta M_2}{M_3}\sin(2\theta_{23})\cos\theta_{12}
\nonumber\\
\beta&\simeq&\frac{m_1+m_3}{2 (m_1-m_3)}
\frac{\delta M_2}{M_3}\sin(2\theta_{23})\sin\theta_{12}\,.
\end{eqnarray}
The parameters $m_\alpha$, $m_\beta$, $m_\gamma$ are of order of the
low energy neutrino masses.
The angles $\alpha$, $\beta$, $\gamma$ are of order $\delta M_i/M_3$
with the exception of degenerate low energy neutrino masses.
In this case $\alpha$ is enhanced by a factor
$m^2/\delta m_{12}^2$, while the other two angles $\beta$ and $\gamma$
have a factor $m^2/\delta m_{13}^2$,
and our approach here is not valid any more because
the three angles can be small only if the degeneracy of the
Dirac neutrino eigenvalues is such that $\delta M_i/M<10^{-5}$.
We notice that there is not any substantial difference for normal
($m_1<m_2<m_3$) or inverted hierarchy ($m_3<m_1<m_2$) of the low
energy neutrino masses, and the only effect is to change the
sign of $\beta$ and $\gamma$ angles.
\\
From eq. (\ref{eq:Our}) we get
\begin{eqnarray}
	\tilde M_D=\Omega^\dagger U_{CKM}^\dagger M_D^\Delta V_M V_\epsilon
\end{eqnarray}
and $\tilde M_D$ can be computed using the expressions in eq. 
(\ref{eq:MDdeg}) and $U_{CKM}$. Notice that in this case the resulting
$\tilde M_D$ strongly depends on the $V_M$ matrix.
\\
For any $V_M$, the heavy neutrino spectrum is degenerate.
However the mixing matrix $V_0^\dagger V_R^\star$ is close to
the $V_M$ matrix.

\subsubsection{Degenerate $m_{low}$}
If the low energy neutrino masses $m_i$ and the Dirac
neutrino eigenvalues are degenerate then we get 
\begin{eqnarray}
V_0^\dagger V_R^\star\frac{1}{M_R^\Delta} V_R^\dagger V_0^\star
&\simeq&
\frac{1}{M_D^\Delta} m_{low}^\Delta\frac{1}{M_D^\Delta}\,.
\end{eqnarray}
In this case the value of $V_M$ plays a marginal role.
The mixing matrix $V_0^\dagger V_R^\star$ is close to
a small rotation in the $(1,3)$ plane and
the heavy neutrino spectrum is degenerate too:
\begin{eqnarray}
M^R_1&=&\frac{m}{M^2}\left(1 - \frac{\delta M_1}{M}
   \left(1+\sqrt{1-\frac{1}{3}\frac{\sqrt{\delta m_{sol}^2}/m}{\delta M_1/M}
   +\frac{\delta m_{sol}^2/m^2}{(\delta M_1/M)^2}}\right)\right)
\nonumber\\
M^R_2&=&\frac{m}{M^2}\left(1-2\frac{\delta M_2}{M}
  +\frac{\sqrt{\delta m_{atm}^2}}{m}\right)\\
M^R_3&=&\frac{m}{M^2}\left(1 - \frac{\delta M_1}{M}
   \left(1-\sqrt{1-\frac{1}{3}\frac{\sqrt{\delta m_{sol}^2}/m}{\delta M_1/M}
   +\frac{\delta m_{sol}^2/m^2}{(\delta M_1/M)^2}}\right)\right)\nonumber\,.
\end{eqnarray}
For any $V_M$ compatible with the experiments,
the heavy neutrino spectrum is almost degenerate.
Moreover the mixing matrix $V_0^\dagger V_R^\star$ is close to
the identity matrix.

\section{Contribution to $l_i\rightarrow l_j \gamma$}\label{sec:muegamma}
Using the result in eq. (\ref{eq:Our}) and the general
eq. (\ref{eq:BR}), we get
\begin{eqnarray}\label{eq:BROur}
BR(l_i\rightarrow l_j\gamma)
     &\propto&
\left|\left(\Omega^\dagger U_{CKM}^\dagger M_D^\Delta 
V L V^\dagger
M_D^\Delta U_{CKM}\Omega \right)_{ij}\right|^2
\end{eqnarray}
where $V=V_0^\dagger V_R^\star$ is the mixing matrix computed with
eq. (\ref{eq:V0VR}). Notice that the $\Omega$ phase differences
$\exp^{i(\phi_i-\phi_j)}$ cancel because
we take the absolute value.
We want to stress here that the result in eq. (\ref{eq:BROur})
depends on the Quark-Lepton complementarity (and the underlying flavor symmetry)
assumption only, and not on the explicit form
of the correlation matrix $V_M$.

At zero approximation we neglect the different
normalizations for different right-handed neutrinos.
We assume $L=\hat L ={\bf 1}\log M_X/M_R$ where $M_R$ is the
common heavy neutrino mass.
The $BR(\mu\rightarrow e\gamma)$ can be rewritten as
\begin{eqnarray}\label{eq:BRmue}
BR(\mu\rightarrow e\gamma)
     &\propto&
\frac{\Gamma(\mu\rightarrow e\nu\nu)}{\Gamma(\mu)}
\frac{\alpha^3}{G_f m_s^8 v_u^4}\tan^2\beta
\left(\frac{3m_0+A_0}{8\pi^2}\right)^2 \hat L^2\,
\nonumber\\&&\quad\quad
\left|\left(U_{CKM}^\dagger (M_D^\Delta)^2 U_{CKM}\right)_{21}\right|^2
\nonumber\\&=&
\frac{\Gamma(\mu\rightarrow e\nu\nu)}{\Gamma(\mu)}
\frac{\alpha^3}{G_f m_s^8 v_u^4}\tan^2\beta
\left(\frac{3m_0+A_0}{8\pi^2}\right)^2 \hat L^2\,
\\&&\quad\quad
\left|(M_2^2-M_1^2)\lambda(1 + O(\lambda^2))
+M_3^2 A^2 (\rho-i\eta) \lambda^5(1 + O(\lambda^6))
\right|^2
\nonumber
\end{eqnarray}
where $\lambda$ is the sine of the Cabibbo angle,
and $A$, $\rho$ and $\eta$ are the other parameters of
the unitary CKM matrix.
For each Dirac neutrino mass we introduced, its first contribution.
Similarly to the process $\mu\rightarrow e\gamma$ we can compute
the contribution to the $\tau$ decays.
For $\tau\rightarrow e\gamma$ we get
\begin{eqnarray}\label{eq:BRtaue}
BR(\tau\rightarrow e\gamma)
     &\propto&
\frac{\Gamma(\tau\rightarrow e\nu\nu)}{\Gamma(\tau)}
\frac{\alpha^3}{G_f m_s^8 v_u^4}\tan^2\beta
\left(\frac{3m_0+A_0}{8\pi^2}\right)^2 \hat L^2\,
\\&&\quad\quad
\left|((1-(\rho-i\eta))M_1^2-M_2^2+M_3^2(\rho-i\eta))
A\lambda^3(1 + O(\lambda^2))
\right|^2\,.
\nonumber\end{eqnarray}
The other $\tau$ decay process that violates the individual
lepton number is such that
\begin{eqnarray}\label{eq:BRtaumu}
BR(\tau\rightarrow \mu\gamma)
     &\propto&
\frac{\Gamma(\tau\rightarrow \mu\nu\nu)}{\Gamma(\tau)}
\frac{\alpha^3}{G_f m_s^8 v_u^4}\tan^2\beta
\left(\frac{3m_0+A_0}{8\pi^2}\right)^2 \hat L^2\,
\\&&\quad\quad
\left|(-M_1^2\lambda^2-M_2^2+M_3^2)
A\lambda^2(1 + O(\lambda^2))
\right|^2\,.
\nonumber
\end{eqnarray}
To understand the main contribution we must make some assumptions
about the hierarchy of the Dirac neutrino masses $M_i$.
Moreover to include the effect of non degeneration for
heavy neutrino masses we must include $V$,
whose form depends also on the hierarchy of the low energy
neutrino masses.

\subsection{Hierarchical $M_D$}
For hierarchical $M_D$ the factor $L$ in eq. (\ref{eq:BROur})
cannot be neglected. If we introduce the full form of $L$ then
the form of $V$
is relevant. Under the assumption of hierarchical $M_D$,
$V$ is close to the identity and we get
\begin{eqnarray}\label{eq:53}
BR(\mu\rightarrow e\gamma)
     &\propto&
\frac{\Gamma(\mu\rightarrow e\nu\nu)}{\Gamma(\mu)}
\frac{\alpha^3}{G_f m_s^8 v_u^4}\tan^2\beta
\left(\frac{3m_0+A_0}{8\pi^2}\right)^2
\\&&\quad
\left|\left(M_2^2 \log\frac{M_X}{M^R_2}-M_1^2\log\frac{M_X}{M^R_3}\right)\lambda
+M_3^2 \log\frac{M_X}{M^R_1} A^2 (\rho-i\eta) \lambda^5
\right|^2\,,
\nonumber
\end{eqnarray}
where we introduced the structure of $L$ to take into account
the hierarchical structure of heavy neutrino masses too.
For example if we assume that 
$$M_1:M_2:M_3\propto m_u:m_b:m_t$$
 at the unification scale, then we obtained in Sec. {\bf\ref{sec:MDdeg}}
that
$$M^R_1:M^R_2:M^R_3\propto m_u^2:m_c^2:m_t^2\,.$$
 For the BR we have
\begin{eqnarray}\label{eq:BRmueHiera}
BR(\mu\rightarrow e\gamma)
     &\propto&
\frac{\Gamma(\mu\rightarrow e\nu\nu)}{\Gamma(\mu)}
\frac{\alpha^3}{G_f m_s^8 v_u^4}\tan^2\beta
\left(\frac{3m_0+A_0}{8\pi^2}\right)^2 \log^2\frac{M_X}{M_3}\,
\\&&\quad\quad
\left(\frac{M_3}{m_t}\right)^4
\left|m_c^2\lambda\log\frac{m_t^2}{m_c^2}
     +m_t^2\log\frac{m_t^2}{m_u^2}A^2 (\rho-i\eta) \lambda^5\right|^2\,.
\nonumber
\end{eqnarray}
%
%\subsection{$\tau\rightarrow e\gamma$ and $\tau\rightarrow \mu\gamma$}
Similarly to the process $\mu\rightarrow e\gamma$ we can compute
the contribution to $\tau\rightarrow e\gamma$
and $\tau\rightarrow \mu\gamma$.
We get
\begin{eqnarray}\label{eq:BRtauedeg}
BR(\tau\rightarrow e\gamma)
     &\propto&
\frac{\Gamma(\tau\rightarrow e\nu\nu)}{\Gamma(\tau)}
\frac{\alpha^3}{G_f m_s^8 v_u^4}\tan^2\beta
\left(\frac{3m_0+A_0}{8\pi^2}\right)^2 \hat L^2\,
\\&&\quad\quad
\left(\frac{M_3}{m_t}\right)^4
\left|m_t^2\log\frac{m_t^2}{m_u^2} A (\rho-i\eta) \lambda^3\right|^2\,,
\nonumber
\end{eqnarray}
where in the last line we used a hierarchical structure for the
Dirac neutrino masses and introduced the structure of $L$.
We observe that $BR(\mu\rightarrow e\gamma)$ is suppressed by
a factor $\lambda^4$ with respect to $BR(\tau\rightarrow e\gamma)$.
\\
The other $\tau$ decay is the least suppressed process that
violates the individual lepton number.
In fact we have
\begin{eqnarray}\label{eq:BRtaumudeg}
BR(\tau\rightarrow \mu\gamma)
     &\propto&
\frac{\Gamma(\tau\rightarrow \mu\nu\nu)}{\Gamma(\tau)}
\frac{\alpha^3}{G_f m_s^8 v_u^4}\tan^2\beta
\left(\frac{3m_0+A_0}{8\pi^2}\right)^2 \hat L^2\,
\\&&\quad\quad
\left(\frac{M_3}{m_t}\right)^4
\left|m_t^2\log\frac{m_t^2}{m_u^2} A \lambda^2\right|^2\,.
\nonumber
\end{eqnarray}
We observe that $BR(\mu\rightarrow e\gamma)$ is in general suppressed by
a factor $\lambda^6$ with respect to $BR(\tau\rightarrow \mu\gamma)$,
and $BR(\tau\rightarrow \mu\gamma)$ by a factor $\lambda^2$.
Our conclusions are equivalent to the one in \cite{Hochmuth:2006xn,Cheung:2005gq},
and also in our analysis it can be a further suppression
of the branching ratios if the leading term in eq. (\ref{eq:53})
cancels.
We can conclude that in this case, for general values of the
SUSY parameters, the expected branching ratios are compatible
with the actual experimental data, and will be observable
only for high value of the low energy neutrino masses and for
particular point in the SUSY parameter space.
However our discussion is more general since in fact we showed that these
results do not depend on the form of the correlation matrix $V_M$.

\subsection{Degenerate $M_D$}
If we assume that the eigenvalues of the Dirac Yukawa matrix
are degenerate, as computed in sec. {\bf\ref{sec:MDdeg}},
we have two cases depending on the degeneration of $m_{low}$.

\subsubsection{Non degenerate $m_{low}$}
For non degenerate $m_{low}$ we have the right-handed
neutrinos with the same hierarchy of the low energy neutrinos,
and $V_0^\dagger V_R^\star$ close to $V_M$.
In this case we get
\begin{eqnarray}
BR(\mu\rightarrow e\gamma)
     &\propto&
\frac{\Gamma(\mu\rightarrow e\nu\nu)}{\Gamma(\mu)}
\frac{\alpha^3}{G_f m_s^8 v_u^4}\tan^2\beta
\left(\frac{3m_0+A_0}{8\pi^2}\right)^2
\nonumber\\&&
\left|
M_1 M_2\log\frac{m_\beta}{m_\alpha}
\cos\alpha_{12} \cos\alpha_{23}\sin\alpha_{12}
\right|^2\,;
\\
BR(\tau\rightarrow e\gamma)
     &\propto&
\frac{\Gamma(\mu\rightarrow e\nu\nu)}{\Gamma(\mu)}
\frac{\alpha^3}{G_f m_s^8 v_u^4}\tan^2\beta
\left(\frac{3m_0+A_0}{8\pi^2}\right)^2
\nonumber\\&&
\left|
M_1 M_3\log\frac{m_\beta}{m_\alpha}
\cos\alpha_{12} \sin\alpha_{23}\sin\alpha_{12}
\right|^2\,;
\\
BR(\tau\rightarrow \mu\gamma)
     &\propto&
\frac{\Gamma(\mu\rightarrow e\nu\nu)}{\Gamma(\mu)}
\frac{\alpha^3}{G_f m_s^8 v_u^4}\tan^2\beta
\left(\frac{3m_0+A_0}{8\pi^2}\right)^2
\nonumber\\&&
\left|
M_2 M_3\cos\alpha_{23}\sin\alpha_{23}
\left(\log\frac{m_\gamma}{m_\alpha}
+\sin^2\alpha_{12}\log\frac{m_\beta}{m_\alpha}\right)
\right|^2\,.
\end{eqnarray}
The ratios among them become of order one
\begin{eqnarray}
\frac{BR(\mu\rightarrow e\gamma)}{BR(\tau\rightarrow e\gamma)}
&\simeq&\tan^2\alpha_{23} \in [0.5,1.4]
\\\nonumber&&\mbox{and}
\\
\frac{BR(\mu\rightarrow e\gamma)}{BR(\tau\rightarrow \mu\gamma)}
&\simeq&
\left|
\frac{\cos\alpha_{12}\sin\alpha_{12}}{
\sin\alpha_{23}\left((\log\frac{m_\gamma}{m_\alpha}/\log\frac{m_\beta}{m_\alpha})
+\sin^2\alpha_{12}\right)}
\right|^2
\end{eqnarray}
We notice that in this case, with respect to the one considered in the previous
section, the value of the branching ratio of $\mu\rightarrow e \gamma$ is bigger
by a factor $\lambda^6$. So we obtain that, despite the fact that this case is
the most promising to extract information on the structure of $V_M$, degenerate
$M_D$ and non degenerate $m_{low}$ is excluded by the experimental data for most
of the SUSY parameters. Naturally one can fine-tuning the SUSY parameter and/or
the neutrino mass parameters in such a way to escape from our general analysis.

\subsubsection{Degenerate $m_{low}$}
If the spectrum of the low energy neutrino is degenerate, then 
the mixing matrix $V_0^\dagger V_R^\star$ becomes close to
the identity. In this case the branching ratios depend on
the common $M_D$ mass and the Cabibbo parameter.
By assuming\footnote{If this relation does not hold then we are in the case
of degenerate $M_D$ and non degenerate $m_{low}$.}
$M_2^2-M_1^2> \lambda^4 M_3^2$ we get
\begin{eqnarray}
BR(\mu\rightarrow e\gamma)
     &\propto&
\left|(M_2^2-M_1^2)\right|^2\lambda^2\,,
\nonumber\\
BR(\tau\rightarrow e\gamma)
     &\propto&
\left|((1-(\rho-i\eta))M_1^2-M_2^2+M_3^2(\rho-i\eta))
\right|^2(A\lambda^3)^2\,,
\nonumber\\
BR(\tau\rightarrow \mu\gamma)
     &\propto&
\left|(M_3^2-M_2^2)\right|^2
(A\lambda^2)^2\,,
\end{eqnarray}
and the ratios among them are
$$
BR(\mu\rightarrow e\gamma):BR(\tau\rightarrow e\gamma):BR(\tau\rightarrow \mu\gamma)=1:\lambda^4:\lambda^2\,.
$$
To compare this case with the case of hierarchical $M_D$
of sec {\bf\ref{sec:hier}}, we
observe that here $BR(\mu\rightarrow e\gamma)$ is the
largest one, while in the other case it is the smallest one. 
Moreover the value of the branching ratios here depends
on the differences $M_i^2-M_j^2$ and they are in general
smaller then in the other case.
For example, if $M_2^2-M_1^2 \simeq \lambda^4 M_3^2$ and
$M_i$ are of order $m_{t}$, we obtain
\begin{eqnarray}
BR(\mu\rightarrow e\gamma)&\propto&
\left(\frac{M_3}{m_t}\right)^4
\left|m_t^4\lambda^5\right|^2\,,
\nonumber\\
BR(\tau\rightarrow e\gamma)&\propto&
\left(\frac{M_3}{m_t}\right)^4
\left|m_t^4\lambda^7\right|^2\,,
\nonumber\\
BR(\tau\rightarrow \mu\gamma)&\propto&
\left(\frac{M_3}{m_t}\right)^4
\left|m_t^4\lambda^6\right|^2\,.
\end{eqnarray}
In this case, not only we cannot extract information on the $V_M$ structure,
but also we have no hope to observe these branching ratios because they
are too small even with respect to the future experimental sensitivities.

\section{Conclusions}\label{sec:conclusion}
We analized the consequences of a non trivial Quark-Lepton complementarity
and a flavor symmetry on $BR(l_i\rightarrow l_j\gamma)$.
The non trivial Quark-Lepton complementarity, together with
the flavor symmetry, states that the correlation matrix $V_M$,
product of the $CKM$ and the  $PMNS$ mixing matrix, is related to
the diagonalization of the Majorana right-handed and Dirac neutrino 
mass matrices.
In this framework we obtained that
$BR(l_i\rightarrow l_j\gamma)$ is related to the $CKM$
mixing matrix and the Dirac neutrino masses.
\\
We have three cases:
\begin{enumerate}
\item{Hierarchical Dirac neutrino eigenvalues 
(very hierarchical right-handed neutrino masses,
$V_0^\dagger V_R^\star\simeq I$) where we get the usual ratios
$$BR(\mu\rightarrow e\gamma):BR(\tau\rightarrow e\gamma):BR(\tau\rightarrow \mu\gamma)
  =\lambda^6:\lambda^4:1 \propto M_3^4 \lambda^4 \hat L\,.$$
This case is the most promising one for a future observation of the
branching ratios. However it will not give us any information
about the structure of the $V_M$ matrix.
}
\item{Degenerate Dirac neutrino eigenvalues, with non degenerate
low energy neutrino masses
(the hierarchy of the right-handed neutrino masses is close to the
one of the low energy spectrum, $V_0^\dagger V_R^\star \simeq V_M$)
where we get
$$BR(\mu\rightarrow e\gamma)= \tan^2\theta_{23}^{V_M}
 BR(\tau\rightarrow e\gamma)
= f(\theta_{12}^{V_M},\theta_{23}^{V_M}) BR(\tau\rightarrow \mu\gamma) \propto M_3^4 \hat L$$
with $f(\theta_{12}^{V_M},\theta_{23}^{V_M})$ of order one.
This case is the only one where the structure of $V_M$ plays a fundamental role
in the determination of the branching ratios. However it is already excluded
for a large part of the SUSY parameters space by the experimental limits.
}
\item{Degenerate Dirac neutrino eigenvalues and low energy neutrino
spectrum (right-handed neutrinos close to degenerate,
$V_0^\dagger V_R^\star\simeq I$) where we have
$$BR(\mu\rightarrow e\gamma):BR(\tau\rightarrow e\gamma):BR(\tau\rightarrow \mu\gamma)
=1:\lambda^4:\lambda^2 \propto M_3^4\lambda^{10} \hat L\,.$$
In this case the branching ratios are too small even with respect to
the future experimental sensitivities.
}
\end{enumerate}
\subsection*{Acknowledgments}
We thank Jo\~ao Pulido for useful discussions
about neutrino physics, and Jorge C. Rom\~ao for enlightening
discussion about flavor violating processes in supersymmetry.
We acknowledge the MEC-INFN grant, and
the Funda\c{c}\~{a}o para a Ci\^{e}ncia e a Tecnologia for the grant SFRH/BPD/25019/2005.

\end{document}